\documentclass[12pt]{iopart}
\usepackage{iopams}
\expandafter\let\csname equation*\endcsname\relax
\expandafter\let\csname endequation*\endcsname\relax
\usepackage{amsmath,amssymb,bm,yfonts}
\usepackage{aas_macros}
\usepackage[pdftex]{hyperref}
\usepackage{cite}
\usepackage[pdftex]{graphicx}
\usepackage[utf8]{inputenc}
\usepackage[normalem]{ulem}

\mathchardef\mhyphen="2D
\usepackage{xcolor}

\makeatletter
\g@addto@macro\@floatboxreset\centering
\makeatother

\begin{document}

\title{Light echos and coherent autocorrelations in a black hole spacetime}

\author{Paul~M.~Chesler$^{2}$,
        Lindy~Blackburn$^{1}$,
	    Sheperd~S.~Doeleman$^{1,2}$,
	    Michael~D.~Johnson$^{1,2}$,
	    James~M.~Moran$^1$,
	    Ramesh~Narayan$^{1,2}$,
	    and Maciek~Wielgus$^{1,2}$
}
	    
\address{$^1$ Black Hole Initiative, Harvard University, Cambridge, MA 02138, USA}
\address{$^2$ Center for Astrophysics $\vert$ Harvard \& Smithsonian, 60 Garden Street, Cambridge, MA 02138, USA}

\vspace{10pt}

\begin{indented}
\item[] \today
\end{indented}

\begin{abstract}
  The Event Horizon Telescope recently produced the first images of a black hole. These images were synthesized by measuring the coherent correlation function of the complex electric field measured at telescopes located across the Earth. This correlation function corresponds to the Fourier transform of the image under the assumption that the source emits spatially incoherent radiation. However, black holes differ from standard astrophysical objects: in the absence of absorption and scattering, an observer sees a series of increasingly demagnified echos of each emitting location. These echos correspond to rays that orbit the black hole one or more times before reaching the observer. This multi-path propagation introduces spatial and temporal correlations into the electric field that encode properties of the black hole, irrespective of intrinsic variability. We explore the coherent temporal autocorrelation function measured at a single telescope. Specifically, we study the simplified toy problem of scalar field correlation functions $\langle \Psi(t) \Psi(0) \rangle$ sourced by fluctuating matter located near a Schwarzschild black hole. We find that the correlation function is peaked at times equal to integer multiples of the photon orbit period; the corresponding power spectral density vanishes like $\lambda/r_{\rm g}$ where $r_{\rm g} = G M / c^{2}$ is the gravitational radius of the black hole and $\lambda$ is the wavelength of radiation observed. For supermassive black holes observed at millimeter wavelengths, the power in echos is suppressed relative to direct emission by $\sim 10^{-13} \lambda_{\rm mm}/M_{6}$, where $\lambda_{\rm mm} = \lambda/(1\,{\rm mm})$ and $M_6 = M/(10^6 M_\odot)$. Consequently, detecting multi-path propagation near a black hole using the coherent electric field autocorrelation is infeasible with current technology.   
\end{abstract}

%
%
%
%
%

\section{Introduction}

LIGO's discovery of binary black hole mergers 
\cite{Abbott:2016blz,
	  Abbott:2016nmj,
	  TheLIGOScientific:2016pea,
	  Abbott:2017vtc,
	  Abbott:2017gyy,
	  LIGOScientific:2020stg} 
and the EHT's first images of the shadow of a supermassive black hole
\cite{Akiyama:2019cqa,
      Akiyama:2019brx,
      Akiyama:2019sww,
      Akiyama:2019bqs,
      Akiyama:2019fyp,
      Akiyama:2019eap} 
provide an unprecedented opportunity to study the near-horizon spacetime geometry of black holes and test General Relativity in extreme conditions. General relativity predicts the existence of 
bound null orbits in the black hole's photon shell \cite{Bardeen_1973,Teo_2003}.  The bound orbits are unstable, meaning light from nearby orbits can escape to infinity and contribute to the black hole's image.  In particular, light propagating along trajectories close to bound orbits produces a sharp feature in the image, the photon ring, with light rays asymptotically close to bound orbits forming the edge of the black hole shadow, the ``critical curve'' \cite{Bardeen_1973,Johnson:2019ljv,Gralla:2019drh,Vincent_2020}.

\begin{figure}[h]
	\includegraphics[trim= 100 300 500 300 ,clip,scale=0.40]{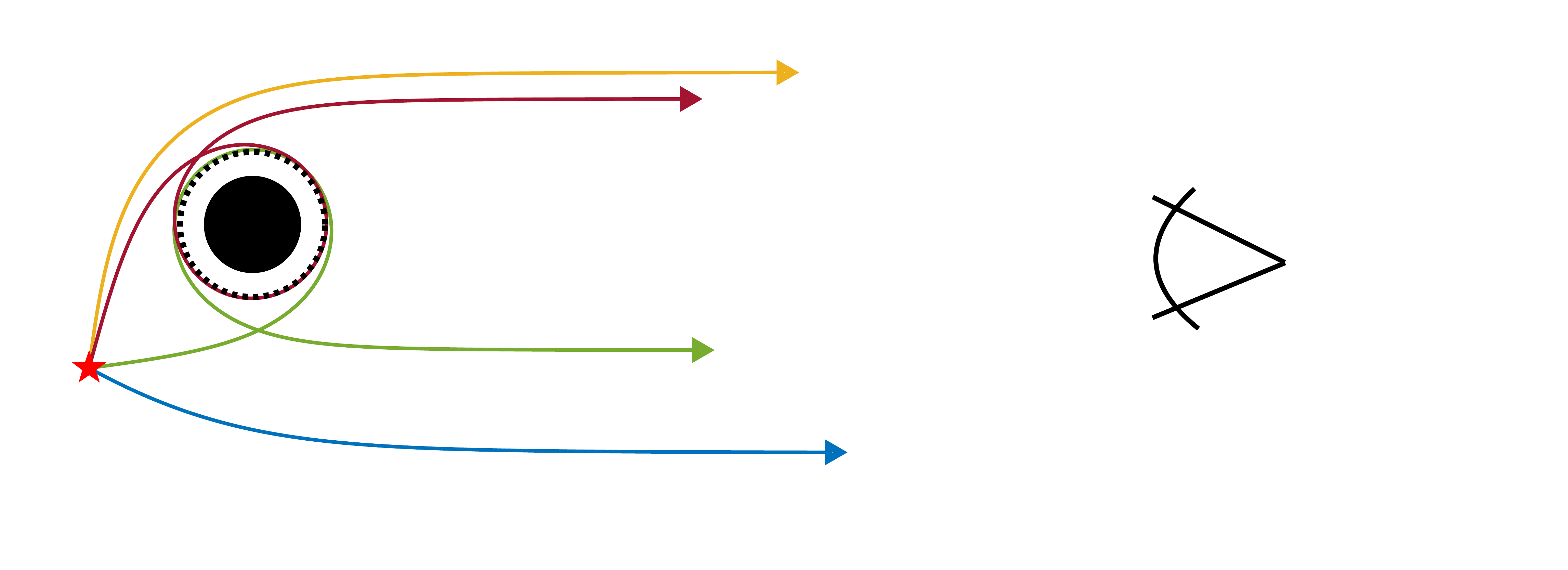}
	\caption{Four light rays shown to illustrate multi-path propagation in the     Schwarzschild spacetime. The solid black disc denotes the black hole while the dashed circle denotes its photon sphere, which is located at $r = 3 M$.
	Each light ray is emitted at the red star and eventually propagates to a distant observer to the right.  The path the light takes depends on the direction of emission.  Light rays can propagate directly to the observer (albeit along lensed trajectories) or can orbit the black hole several times before escaping, with the associated orbits lying close to the photon sphere.  Light rays that orbit the black hole arrive at the observer delayed relative to direct trajectories.
	}
	\label{fig:geos}
\end{figure}

While the photon shell can manifest itself in black hole images, it also imparts time-dependent signatures. 
Namely, suppose a burst of light is emitted just outside the photon shell, as depicted in Fig.~\ref{fig:geos}.  Light from the burst can take multiple paths to a distant observer, including a direct path (blue), a partial orbit (yellow), or complete orbits (green or maroon).  A distant observer would therefore see a primary burst from the direct light, a delayed lensed burst coming from light that partially orbited the black hole, and a series of echoed bursts from light that orbited the black hole one or more times \cite{Moriyama_2019}. The echoed bursts are approximately separated in time by multiples of the photon orbit period and are exponentially attenuated in amplitude due to successive demagnification \cite{Johnson:2019ljv,Darwin_1959,Luminet_1979}. 

Due to the fact that accretion flows around supermassive black holes are continuously emitting light, it is natural to look for signs of multi-path propagation and light echos 
in correlation functions. A reasonable expectation is the correlation functions should contain structure at integer multiples of the photon orbit period. To compare with observations, there are two fundamental correlation functions to consider. The first is the correlation of the quasimonochromatic and complex scalar electric field measured at an observing frequency $\nu$, $\langle E_\nu(t) E_\nu^\ast(t') \rangle$, which is related to the power spectral density by a Fourier transform. The second is the correlation of the flux density $I_\nu \sim \langle \left| E_\nu \right|^2 \rangle$ (i.e., the ``light-curve''), $\langle I_\nu(t) I_\nu(t') \rangle$ \cite{TMS}. While many astrophysical processes can introduce correlation structure in light curves, astrophysical sources emit spatially and temporally incoherent radiation, giving a temporally incoherent signal for a distant observer: $\langle E_\nu(t) E_\nu^\ast(t') \rangle \sim \delta_{\Delta \nu}(t-t')$, where the delta response has a width comparable to the inverse bandwidth $1/\Delta \nu$.   
Thus, multi-path propagation from the photon shell of a black hole imprints unmistakable signatures in the electric field autocorrelation structure, even for a static source.\footnote{While some physical effects, such as scattering, introduce non-trivial correlation structure in the electric field, the correlations introduced by multi-path propagation near a black hole would be unmistakable, appearing as strongly delayed and discrete peaks above a vanishing background.} 
Measuring non-zero autocorrelation at a large delay $\Delta t \gg 1/\Delta \nu$ would then demonstrate that the received light had executed wraps around a compact object, demonstrating that the object's mass lies within its photon orbit, and measuring the delay spectrum of the object would give new constraints on the underlying spacetime metric. 

In this paper, we explore the expected autocorrelation signal from a black hole encoded in the electric field. In particular, millimeter telescopes routinely record the complex field when participating in very long baseline interferometry (VLBI) experiments, such as the EHT, as the \emph{spatial} correlations in this field are used to produce images. 
We instead propose to study the \emph{temporal} correlations in this field, which could be done by analyzing recorded baseband data at a single telescope. To derive estimates for the expected correlation structure, we wish to study light echos in the simplest possible setting.  To this end, instead of studying electrodynamics sourced by fluctuating electric currents, we study a toy model composed of a massless scalar field $\Psi$ sourced by a stochastic field $\rho$ localized near the black hole. 
Furthermore, since the photon shell of Kerr black holes contains a distribution of photon orbit periods whose observation depends on one's inclination \cite{Gralla:2019ceu,Gralla:2019drh}, we choose to restrict our attention to Schwarzschild black holes, where there is a single photon orbit period 
\begin{equation}
\label{eq:T}
T = 2 \pi r_\gamma (1-2M/r_\gamma)^{-1/2} = 6\pi \sqrt{3}M,
\end{equation}
associated with the photon sphere located at the radius $r_\gamma = 3M$.
 
We numerically construct scalar field correlation functions and find that they are peaked at integer multiples of $T$. 
We also study the power spectral density, which encodes the power in echos at a given angular frequency $\omega = 2\pi \nu$. 
We find that the power in echos decays like $1/(M \omega)$, where $M$ is the mass of the black hole.  The $1/(M \omega)$ decay is a consequence of cancellations from different emission points and makes 
observing the signature of echos in field correlators challenging, if not impossible, for supermassive black holes. 

An outline of our paper is as follows.  In Sec.~\ref{sec:setup} we present the setup of the problem we wish to solve.  In Sec.~\ref{sec:num} we outline our numerical procedures.  In Sec.~\ref{sec:results} we present our results, and in Sec.~\ref{sec:disc} we discuss our results within the framework of geometric optics.

\section{Setup}
\vspace{20pt}

\label{sec:setup}
We work in geometric units where $G = c = 1$ and employ Schwarzschild coordinates, where the metric takes the form
\begin{equation}
ds^2 = -f dt^2 + \frac{1}{f} dr^2 + r^2 [d \theta^2 + \sin^2 \theta d\phi^2], \ \ \ f = 1 - \frac{2 M}{r}
\end{equation}
for a black hole of mass $M$.  The equation of motion for the scalar field $\Psi$ is just the wave equation,
\begin{equation}
\label{eq:eqm}
-\nabla^2 \Psi = \rho.
\end{equation}
We assume that the source $\rho$ is a spatially and temporally incoherent random field,
\begin{equation}
\label{eq:stats}
\langle \rho(t,\bm r) \rho(t',\bm r') \rangle = \frac{\chi(r)}{\sqrt{g}} \delta(t - t') \delta^3(\bm r - \bm r'),
\end{equation}
for some radial profile function $\chi(r) \geq 0$, which characterizes the strength of fluctuations in $\rho$.  We shall assume that $\chi(r)$ is localized near the black hole.
  
To study echos we employ the correlation function,
\begin{equation}
\label{eq:correlator}
C(t,r) \equiv \langle \Psi(t,\bm r) \Psi(0,\bm r) \rangle,
\end{equation}
and the power spectral density,
\begin{equation}
\widetilde C(\omega,r) \equiv \langle | \hat \Psi(\omega,\bm r)|^2 \rangle,
\end{equation}
where the mode amplitude $\hat \Psi(\omega,\bm r)$ is given by the windowed 
Fourier transform, 
\begin{equation}
\hat \Psi(\omega,\bm r) \equiv \frac{1}{\sqrt{t_{\rm win}}} \int_{-t_{\rm win}/2}^{t_{\rm win}/2} dt \, \Psi(t,\bm r) e^{i \omega t},
\end{equation}
with $t_{\rm win}$ the window duration. $C(t,r)$ measures how signals separated by time $t$ are correlated whereas $\widetilde C(\omega,r)$ measures the amplitude of modes with frequency $\omega$.  We shall consider the limit $t_{\rm win} \to \infty$, in which case the correlation function and the power spectral density are related by Fourier transform,
\begin{equation}
\widetilde C(\omega,r) = \int dt \, C(t,r) e^{i \omega t}.
\end{equation}
Additionally, we shall consider the limit $r \to \infty$, meaning the limit where observations are made arbitrarily far from the black hole.

The equation of motion (\ref{eq:eqm}) is solved by
\begin{equation}
\label{eq:sol}
\Psi(t,\bm r) = \int \sqrt{-g} \,dt' d^3 r' \, G(t-t',\bm r,\bm r') \rho(t',\bm r'),
\end{equation}
where the retarded Green's function $G(t,\bm r, \bm r')$ satisfies
\begin{equation}
\label{eq:greenseqm}
-\nabla^2 G(t,\bm r,\bm r') = \frac{1}{\sqrt{-g}} \delta(t) \delta^3(\bm r - \bm r').
\end{equation}
From the solution (\ref{eq:sol}) and the statistics (\ref{eq:stats}), it follows that
the correlation function (\ref{eq:correlator}) is given by 
\begin{align}
\label{eq:C}
C(t,r) = \int \sqrt{-g} \, dt' d^3 r' \, G(t-t',\bm r,\bm r') G(-t',\bm r,\bm r') \chi(r').
\end{align}
Taking the Fourier transform then yields
\begin{equation}
\label{eq:Ctilde}
\widetilde C(\omega,r) = \int \sqrt{-g} \, d^3 r' \, |\widetilde G(\omega,\bm r,\bm r')|^2 \chi(r'),
\end{equation}
where 
\begin{equation}
\widetilde G(\omega,\bm r,\bm r') = \int dt \, G(t,\bm r,\bm r') e^{i \omega t},
\end{equation}
is the frequency space Green's function.  The problem of computing $\widetilde C(\omega,r)$ and hence $C(t,r)$ therefore reduces to computing $\widetilde G(\omega,\bm r,\bm r')$.  
   
Rotational invariance of the Schwarzschild geometry implies $\widetilde G$ can be expanded in a spherical harmonic expansion in angles,
\begin{equation}
\label{eq:Gdecomp}
\widetilde G(\omega,\bm r,\bm r') = \sum_{\ell m}  y_{\ell m}(\hat r) y^*_{\ell m}(\hat r') \mathcal G_{\ell}(r,r'),
\end{equation}
where $y_{\ell m}$ are spherical harmonics and $\hat r = \frac{\bm r}{r}$.   
The Green's function equation of motion (\ref{eq:greenseqm}) requires
the radial function $\mathcal G_\ell(r,r')$ satisfy the ODE,
\begin{equation}
\label{eq:Gell}
\left [ \frac{\partial}{\partial r} r^2 f \frac{\partial}{\partial r} + \frac{r^2 \omega^2 - \ell(\ell+1) f)}{ f} \right] \mathcal G_{\ell}(r,r') =  \delta(r - r').
\end{equation}
Substituting (\ref{eq:Gdecomp}) into (\ref{eq:Ctilde}) and using the orthogonality of the spherical harmonics as well as the spherical harmonic addition theorem,
\begin{equation}
\sum_{m} | y_{\ell m}(\hat r)|^2 = \frac{2 \ell + 1}{4 \pi},
\end{equation}
one obtains
\begin{equation}
\label{eq:Cintermediate}
\widetilde C(\omega,r) = \frac{1}{4 \pi} \sum_\ell (2 \ell + 1) \int r'^2 d r' \chi(r')   |\mathcal G_\ell(r,r')|^2.
\end{equation}

We now turn to constructing $\mathcal G_\ell (r,r')$.  When $r \neq r'$ the r.h.s.
of Eq.~(\ref{eq:Gell}) vanishes.  It follows that when $r \neq r'$ the function $\mathcal G_{\ell}(r,r')$ must be a linear combination of solutions $h_\ell^\pm(r)$  
to the homogeneous equation of motion,
\begin{equation}
\label{eq:homoeq}
\left [ \frac{\partial}{\partial r} r^2 f \frac{\partial}{\partial r} + \frac{r^2 \omega^2 - \ell(\ell+1) f)}{ f} \right] h^\pm_{\ell}(r) = 0.
\end{equation}
The function $h^-_\ell$ satisfies incoming boundary conditions at the horizon, so the black hole doesn't radiate, whereas $h^+_\ell$ satisfies outgoing boundary conditions at $r = \infty$.  Explicitly, 
\begin{align}
\label{eq:bc}
h^+_\ell(r) \to \frac{e^{i \omega r}}{r} \ {\rm as} \ r \to \infty \ \ {\rm and}\ \ 
h^-_\ell(r) \to e^{-2 i \omega M \log f(r)} \ \ {\rm as} \ \ r \to 2 M.
\end{align}
The appropriate linear combination of $h^\pm_\ell$ is fixed by the requirement that  
$\mathcal G_\ell(r,r')$ is continuous across $r = r'$, but 
has a discontinuous first derivative, which is necessary to obtain the delta function of the r.h.s. of Eq.~(\ref{eq:Gell}).  A short exercise shows
\begin{equation}
\label{eq:Gsol}
\mathcal G_\ell(r,r') = \frac{1}{r'^2 f(r') W(r')}
\begin{cases}
h^-_\ell(r') h^{+}_\ell(r), & r > r', \\
h^+_\ell(r') h^{-}_\ell(r), & r < r',
\end{cases}
\end{equation}
where $W$ is the Wronksian of $h^\pm_\ell$, 
\begin{equation}
W = h^-_\ell \frac{d h^+_\ell}{dr} - h^+_\ell \frac{d h^-_\ell}{dr}.
\end{equation}

With the solution (\ref{eq:Gsol}) and the boundary condition (\ref{eq:bc}), in the large $r$ limit Eq.~(\ref{eq:Cintermediate}) becomes
\begin{equation}
\label{eq:Cnice}
\widetilde C(\omega,r)=\frac{1}{4 \pi r^2} \sum_\ell (2 \ell + 1) \int  r'^2 d r' \chi(r')  \, \left |\frac{h^-_\ell(r')}{ r'^2 f(r') W(r')} \right |^2.
\end{equation}
Note that the only $r$ dependence in (\ref{eq:Cnice}) appears in the $\frac{1}{ 4 \pi r^2}$ prefactor.  In contrast, the integration in (\ref{eq:Cnice}) only involves quantities evaluated near the black hole.  This decomposition is convenient for numerical evaluation.

\begin{figure}[h]
	\includegraphics[trim= 0 0 0 0 ,clip,scale=0.15]{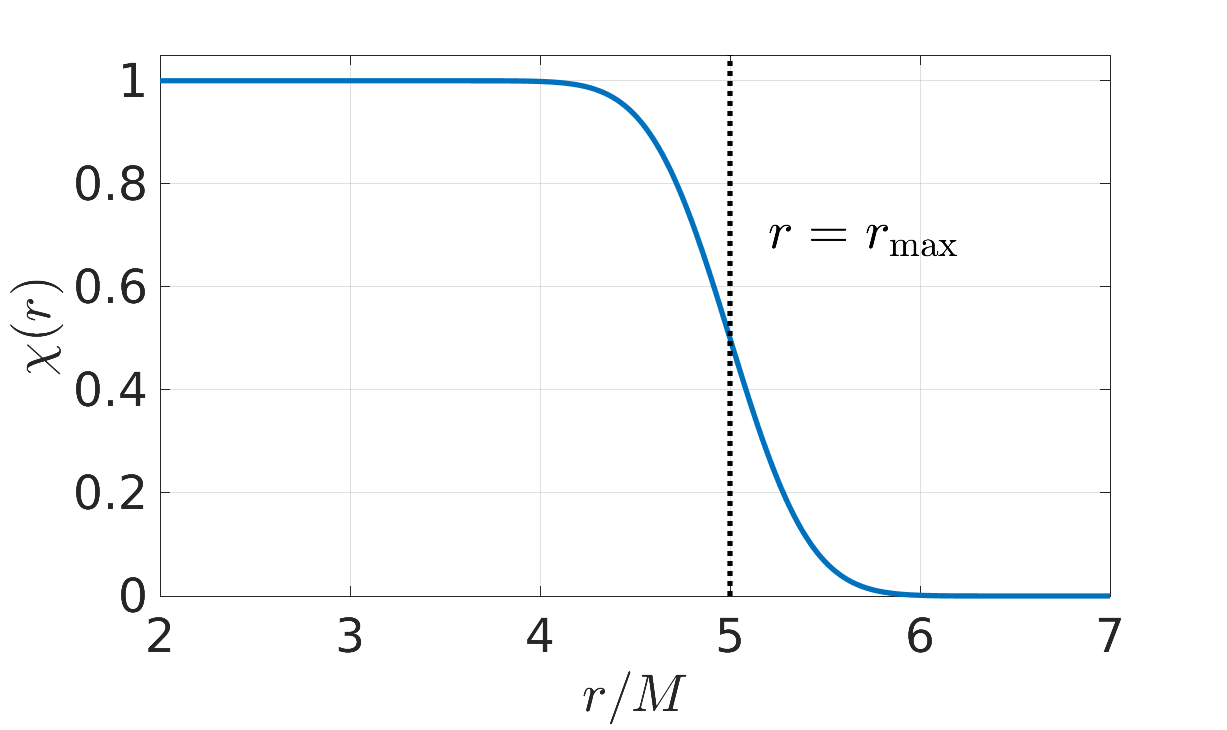}
	\caption{The function $\chi(r)$, given by Eq.~(\ref{eq:Pi}), with $r_{\rm max} = 5 M$.}
	\label{fig:chi}
\end{figure}

\section{Numerics}
\label{sec:num}

For simplicity, in our numerics below we consider 
\begin{equation}
\label{eq:Pi}
\chi(r) = \frac{1}{2} \left [ 1 + {\rm erf}\left (\frac{r - r_{\rm max}}{\sqrt{2} \Delta r} \right ) \right ],
\end{equation}
where ${\rm erf}(z)$ is the error function.  $\chi(r)$ is a smoothed step function, approaching unity when $r - r_{\rm max} \ll - \Delta r$ and exponentially small when $r - r_{\rm max} \gg \Delta r$.  We choose smoothing width $\Delta r = \frac{M}{3}$ and maximum radii $r_{\rm max} = 4M,5M, 6M$. The function $\chi(r)$ is plotted in Fig.~\ref{fig:chi} for $r_{\rm max} = 5 M$.

We determine the functions $h^\pm_\ell(r)$ numerically.  To this end it is useful to define 
\begin{equation}
H_\ell^\pm(r) \equiv h^\pm_\ell(r) \exp\left [\mp i \omega \int dr \frac{1}{f(r)} \right ].
\end{equation}
The functions $H_\ell^\pm(r)$ are just ingoing and outgoing wave functions in ingoing and outgoing Bondi-Sachs coordinates.  In particular the boundary conditions (\ref{eq:bc}) imply that near the horizon $H^-_\ell(r) \sim {\rm const.}$ and at large distances $H^+_\ell(r) \sim 1/r.$  
Removing the oscillatory behavior of $h^+_\ell$ at large distances and that of $h^-_\ell$ near the horizon hastens the numerical computation of these functions.

\begin{figure}[h]
	\includegraphics[trim= 0 0 0 0 ,clip,scale=0.15]{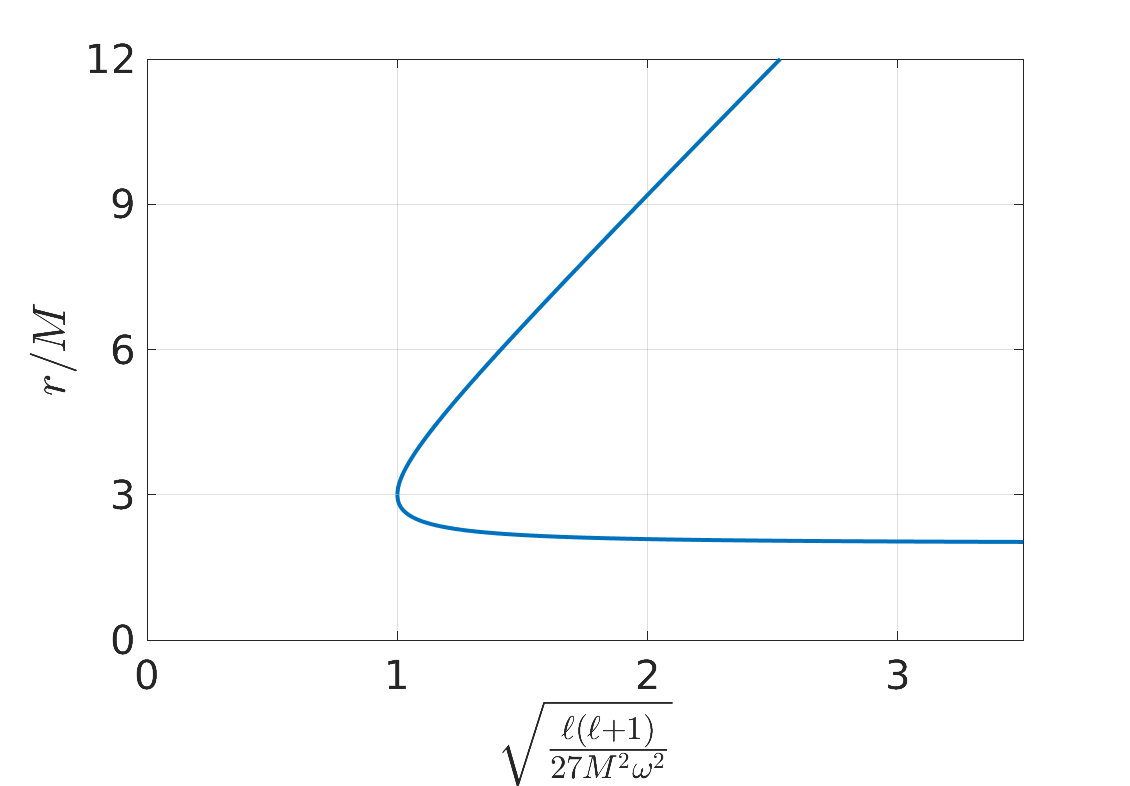}
	\caption{The location of turning points in the ODE (\ref{eq:homoeq}), given by 
		Eq.~(\ref{eq:turningpoints}).  At radii inside the outermost turning point, the summand in Eq.~(\ref{eq:Cnice}) is exponentially small.  
	}
	\label{fig:turningpoints}
\end{figure}

To compute $H^+_\ell$ we integrate in from $r = \infty$.  We accomplish this by breaking the computational domain into two pieces: $r \in (r_{+},\infty)$ and $r\in(2 M,r_{+})$ for some $r_{+}$.  In the outer domain we employ an inverse radial coordinate $z \equiv \frac{1}{r}$ and solve the equation of motion (\ref{eq:homoeq}) using pseudospectral methods (see e.g. \cite{boyd01,Chesler:2013lia}).  With the outer solution constructed, we then integrate inwards from $r = r_{+}$ to $r = 2 M$ using a 4$^{\rm th}$ order Runge-Kutta solver, with boundary data determined by the outer solution.  We choose $r_{+} = 40 M$.

We follow a similar procedure to compute $H^-_\ell.$  We break the computational domain 
up into two segments: $r \in (2 M,r_-)$ and $r \in (r_-,r_{+})$.  In the inner domain we solve the equations of motion using pseudospectral methods.  
With the inner solution constructed, we then integrate outwards from $r = r_-$ using a 4$^{\rm th}$ order Runge-Kutta solver, with boundary data determined by the interior solution.  We choose $r_{-} = 2(1 + 10^{-4}) M$.  

The homogeneous equation of motion (\ref{eq:homoeq}) contains ``turning points" at radii satisfying  
\begin{equation}
\label{eq:turningpoints}
r^2 \omega^2 - \ell (\ell+1) f(r) = 0.
\end{equation}
In Fig.~\ref{fig:turningpoints} we plot the location of the turning points.  With the exception of a single turning point at $r = 3 M$ when $\omega^2 = \ell(\ell+1)/27 M^2$, the turning points always come in pairs.  A WKB analysis demonstrates that inside the outer turning point $h^{-}_\ell$ decreases exponentially with decreasing $r$.  Since the outer turning point diverges like $\ell/\omega$, and $\chi(r)$ is localized at $r \lesssim r_{\rm max}$, it follows that the summand in Eq.~(\ref{eq:Cnice}) becomes exponentially small as $\ell \to \infty$.  Correspondingly, in our numerics we truncate the sum over $\ell$ at 
\begin{equation}
\ell_{\rm max} = {\rm max}(16 M \omega,30).
\end{equation}
We have verified that our results below are insensitive to this angular momentum cutoff.  For example, decreasing the cutoff by 20\% produces differences which are smaller than the line width of all the plots presented below.

\section{Results}
\label{sec:results}

\begin{figure}[h]
	\includegraphics[trim= 100 0 0 0 ,clip,scale=0.14]{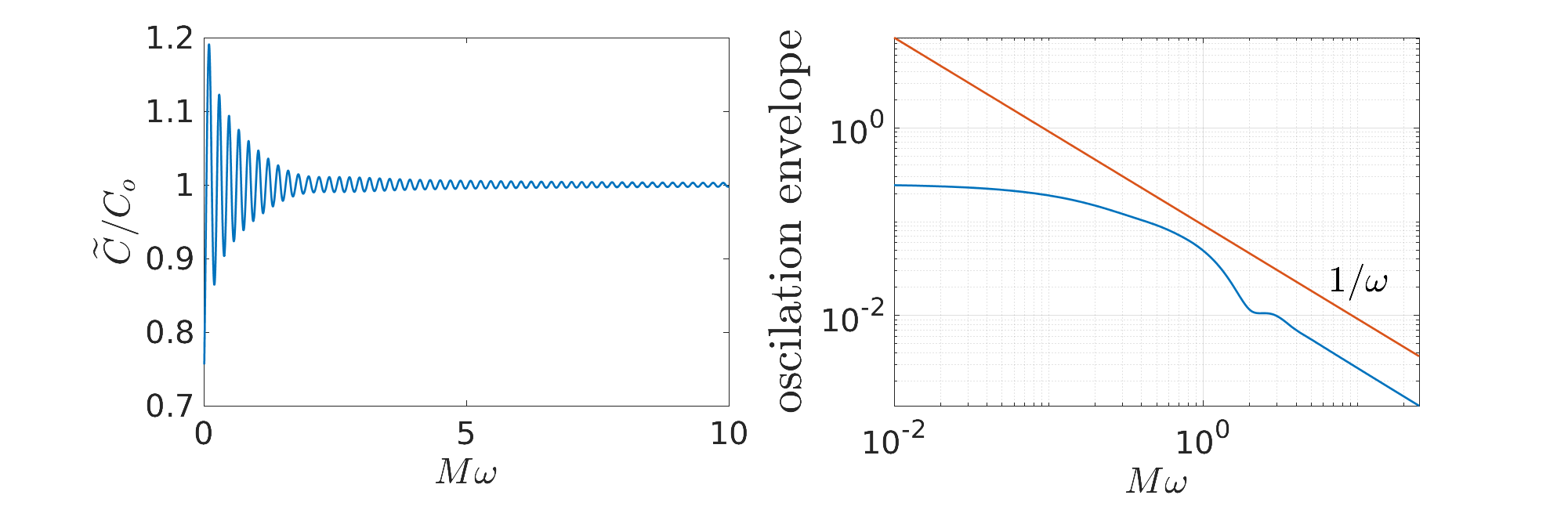}
	\caption{Left: the power spectral density $\widetilde C(\omega,r)$ with $r_{\rm max} = 5M.$  
	At large frequencies the spectral density approaches a constant $C_o$, which we have normalized the plot by.  The oscillations arise from multi-path propagation and roughly have phase $\omega T$, with $T$ the photon orbit period.  Right: the envelope of the oscillations.  At large frequencies the envelope decays like $\frac{1}{\omega}$.
	}
	\label{fig:FTC}
\end{figure}

In the left panel of Fig.~\ref{fig:FTC}, we plot the power spectral density $\widetilde C(\omega,r)$ for $r_{\rm max} = 5 M$.  At large frequencies $C(\omega,r)$ approaches a constant $C_o$, which we have normalized all our plots by.  Also seen in the plots are oscillations, which are most prevalent at low frequencies.  As we shall elaborate on below in the Discussion section, the constant offset $C_o$ arises from direct light propagation from the source to the observer, whereas the oscillations arise from multi-path propagation -- light echos.  The phase of the oscillations is roughly $\omega T$, where $T$ is the black hole's photon orbit period (Eq.~\ref{eq:T}).  In the right panel of Fig.~\ref{fig:FTC} we plot the envelope of the oscillations on a logarithmic scale.%
\footnote{We use Matlab's \textit{envelope} function to compute the envelope.}  Also shown for comparison is the line $\frac{1}{\omega}$.
As is evident from the figure, our numerics are consistent with the envelope decaying like $\frac{1}{\omega}$ as $\omega \to \infty$.

To compute the real time correlator $C(t,r)$, we first construct the difference $(\widetilde C - C_o)$. Subtracting $C_o$ results in a Fourier integrand which decays like $1/\omega$ at large $\omega$, and only changes the resulting Fourier transform by a delta function, $C_o \delta(t)$.  To ameliorate potential logarithmic divergences arising from the $1/\omega$ decay, we also multiply by a window function $\mathcal W(\omega)$, which is identical in functional form to Eq.~(\ref{eq:Pi}) with the replacements $r \to \omega$, $r_{\rm max} \to \omega_{\rm max}$ and $\Delta r \to \Delta \omega$.  We employ maximum frequency $\omega_{\rm max} = 15/M$ and width $\Delta \omega =  2.5/M$. We then Fourier transform $(\widetilde C(\omega,r) - C_o) \mathcal W(\omega)$.  Note that employing a window function means our plots of $C(t,r)$ below lack resolution over temporal scales $\lesssim \frac{1}{\omega_{\rm max}} = \frac{M}{15}$.

\begin{figure}[t]
	\includegraphics[trim= 250 0 0 0 ,clip,scale=0.14]{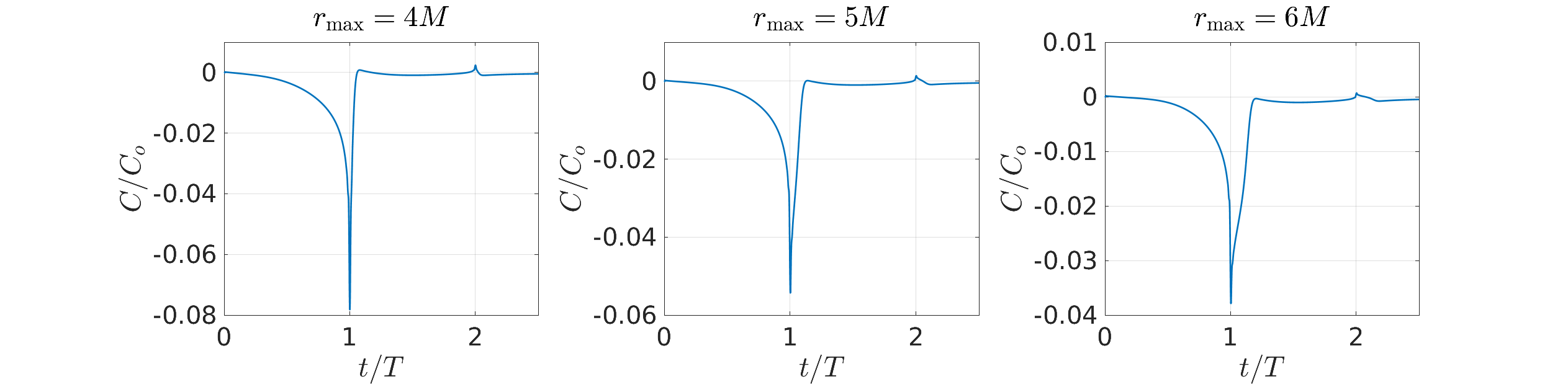}
	\caption{The correlation function $C(t,r)$ for $r_{\rm max} = 4 M$ (left), $5 M$ (middle), and $6 M$ (right).  In all plots there are peaks at integer multiples of the photon orbit period $T$, which alternate in sign.  These peaks are signatures of light echos in the Schwarzschild spacetime.
	}
	\label{fig:realtimecor}
\end{figure}

In Fig.~\ref{fig:realtimecor} we plot $C(t,r)/C_o$ (minus the delta function at $t = 0$) for $r_{\rm max} = 4M$ (left) $r_{\rm max} = 5 M$ (middle) and $r_{\rm max} = 6 M$ (right).  In all plots $C(t,r)$ is generally nonzero at all times.  The most striking feature in the plots is the existence of peaks at $t = T$ and $t = 2 T$. The peaks have alternating signs, with those at $t = 2 T$ having smaller amplitude than those at $t = T$.  Notice that the peaks broaden as $r_{\rm max}$ is increased.  As we elaborate on below, these peaks are signatures of light echos in the Schwarzschild spacetime.

\section{Discussion}
\label{sec:disc}

At wavelengths small compared to the local curvature scale, which for the Schwarzschild spacetime is set by $M$, solutions to the scalar wave equation (\ref{eq:eqm}) are governed by geometric optics (for a pedagogical review see \cite{1973grav.book.....M}).  It follows that the 
high frequency behavior of $\widetilde C(\omega,r)$ and the short-time structure of the peaks in $C(t,r)$ are governed by geometric optics.  

At frequencies $\omega \gg 1/M$, the Green's function $\widetilde G$ can be factored into a slowly varying amplitude and a rapidly varying phase \cite{1973grav.book.....M}, 
\begin{equation}
\label{eq:Goptics}
\widetilde G(\omega,\bm r,\bm r') = \sum_{p} L_p(\bm r,\bm r')e^{i \omega \tau_{p}(\bm r,\bm r')}.
\end{equation}
The sum is over all null geodesics $p$ which connect the emission point $\bm r'$ to the observation point $\bm r$, with $\tau_p(\bm r,\bm r')$ the associated retarded time.  For a black hole geometry there are infinitely many such geodesics, since a geodesic can encircle the black hole an arbitrary number of times before escaping to the observation point $\bm r$.  Examples of such geodesics are shown in Fig.~\ref{fig:geos}. 
The slowly varying amplitudes $L_p(\bm r,\bm r')$ encode the expansion of null geodesics emanating from $\bm r'$ (i.e. demagnification). Scaling relations for $L_p$ can easily be obtained by matching onto the dispersion relation of quasinormal modes (see e.g. \cite{Yang:2012he}). At high angular momentum $\ell$, the longest lived  quasinormal modes have frequencies 
\cite{Schutz:1985km}
\begin{equation}
\label{eq:QNM}
\omega_{\rm QNM} = \pm \frac{2 \pi}{T} \left (\ell + \frac{1}{2} \right ) - \frac{i \pi}{T}.
\end{equation}
Notice that this is simply the large $\ell$ limit of the dispersion relation of a damped two dimensional wave equation
on a sphere.  The dampening reflects the fact that photon orbits are unstable: up to a factor of $-1/T$, the imaginary part of $\omega_{\rm QNM}$ coincides with the  Lyapunov exponent of the photon orbit geodesics \cite{Luminet_1979}.
Owing to the fact that $e^{-i \omega_{\rm QNM} (t + T) } = - e^{-\pi} e^{-i \omega_{\rm QNM} t}$,  it follows that 
\begin{equation}
\label{eq:ampscaling}
L_p \sim (-1)^n e^{-\pi n},
\end{equation}
where $n$ is the number of times the associated light ray orbits the black hole.

Substituting Eq.~(\ref{eq:Goptics}) into Eq.~(\ref{eq:Ctilde}), the resulting power spectral density reads,
\begin{equation}
\widetilde C(\omega, r) = \widetilde C_{\rm direct}(\omega, r) + \widetilde C_{\rm multi\mhyphen path}(\omega, r),
\end{equation}
where 
\begin{equation}
\widetilde C_{\rm direct}(\omega, r) = \sum_{p} \int \sqrt{-g}\, d^3 r'  \chi(r') L_p(\bm r,\bm r')^2,
\end{equation}
and
\begin{equation}
\label{eq:Cecho}
\widetilde C_{\rm multi\mhyphen path}(\omega, r) = \sum_{p\neq p'} \int \sqrt{-g} \, d^3 r'  \chi(r') L_p(\bm r,\bm r') L_{p'}(\bm r,\bm r') e^{i \omega \tau_{pp'}(\bm r,\bm r')},
\end{equation}
with $\tau_{pp'} \equiv \tau_p - \tau_{p'}$ the relative propagation time lag between geodesics $p$ and $p'$.

The direct contribution, $\widetilde C_{\rm direct}$, is independent of $\omega$, meaning $C_o = \widetilde C_{\rm direct}$.  In contrast, the multi-path contribution, $\widetilde C_{\rm multi\mhyphen path}$, oscillates and decays as $\omega$ increases. 
It turns out that the decay envelope scales like
\begin{equation}
\label{eq:asymptotics}
\widetilde C_{\rm multi\mhyphen path}(\omega, r) \sim \frac{1}{M \omega},
\end{equation}
which is consistent with the high frequency limit of $\widetilde C$ shown in Fig.~\ref{fig:FTC}.  

To understand the scaling (\ref{eq:asymptotics}), first consider geodesics which encircle the black hole at most order 1 time.  The time delay $\tau_{pp'}(\bm r,\bm r')$ varies by order $M$ as the emission point $\bm r'$ is varied. In the limit $\omega \gg 1/M$, the phase $\omega \tau_{pp'}$ therefore varies rapidly, leading to cancellations from different emission points.  The integration is therefore dominated by regions near emission points where the phase is stationary, $\frac{\partial \tau_{pp'}}{\partial r'_i} = 0.$ 
In fact cylindrical symmetry dictates that the stationary points must form rings. Since the second derivatives of $\tau_{pp'}$ evaluated on a ring must be order $1/M$, it follows that  
cancellations begin to occur at distances $\sim (M \omega)^{-1/2}$ from the rings.  Cylindrical symmetry
then means that the integral in (\ref{eq:Cecho}) vanishes like $\frac{1}{M \omega}$. 
Next consider pairs of geodesics which encircle the black hole $n$ and $n+m$ times, with
\begin{equation}
\label{eq:largen}
n \sim \frac{1}{2 \pi} \log M \omega.
\end{equation}
The geodesic equation implies  
\begin{equation}
\label{eq:taupp}
\tau_{pp'} = \pm  m T [ 1 + O(e^{-2 \pi n})].
\end{equation}
For such geodesics it follows that the phase $\omega \tau_{pp'}$ varies by order 1 as the emission point is varied. However, Eqs.~(\ref{eq:ampscaling}) and (\ref{eq:largen}) imply $L_p L_{p'} \sim e^{-2 \pi n} \sim \frac{1}{M \omega}$.  Hence such geodesics also yield contributions to $\widetilde C_{\rm multi\mhyphen path}$ which decay like $\frac{1}{M \omega}$.

We now turn to $C(t,r)$.  Eq.~(\ref{eq:Cecho}) Fourier transforms to 
\begin{equation}
\label{eq:Cecho2}
C_{\rm multi\mhyphen path}(t, r) = \sum_{p\neq p'} \int \sqrt{-g} \, d^3 r'  \chi(r') L_p(\bm r,\bm r') L_{p'}(\bm r,\bm r') \delta(t- \tau_{pp'}(\bm r,\bm r')),
\end{equation}
This is generically nonzero for all $t$, since $\tau_{pp'}$ varies continuously from $0$ to $\infty$.  However, for pairs of geodesics which encircle the black hole $0$ and $n$ times, $\tau_{pp'} \approx \pm n T$.  Correspondingly, $C(t,r)$ should be peaked at integer multiples of the photon orbit period, with exponentially decreasing amplitude $\sim  e^{-n \pi}$ and alternating sign $(-1)^n$, which is in qualitative agreement with Fig.~\ref{fig:realtimecor}.
Moreover, emission from points farther from the black hole increases the delay time, since
it takes longer for light to propagate from the emission point to the photon sphere.  This means the widths of the peaks should be broader as $r_{\rm max}$ is increased, just as observed in Fig.~\ref{fig:realtimecor}.

The fact that the power in echos decays like $\frac{1}{M \omega}$ makes observing field correlations challenging.  For example, at an observing wavelength of $\lambda = 1\,{\rm mm}$ and for Sgr~A$^*$, whose mass is $M \approx 4 \times 10^6 M_{\odot}$ and total flux density is $F \approx 2\,{\rm Jy}$, we have a peak non-zero correlation of 
\begin{equation}
\widetilde C_{\rm multi\mhyphen path}(\omega) \approx \frac{F}{M \omega} \sim 10^{-14}\,{\rm Jy},
\end{equation}
indicating that the echo power in field correlators is minuscule relative to direct light. 
This is a consequence of the fact that field correlations are sensitive to phase information and susceptible to cancellations, while the total flux density is the incoherent sum of power throughout the source.  Simply put, echos manifest themselves most strongly in $C(t,r)$ at wavelengths on the order of the horizon radius. 

For observations of supermassive black holes, a better option may be to consider correlations in flux density, which is not sensitive to coherent destructive interference.  The accretion flow around Sgr A$^*$ is highly variable, with macroscopic fluctuations occurring over horizon scales \cite{Baganoff_2001,Johnson_2015,Gravity_2018,Witzel_2020}. 
These fluctuations -- and their echos -- should manifest themselves in light curves of flux density \cite{Broderick_2006,Moriyama_2019,Tiede2020,Hadar_2020,Wong_2020}, although these measurements are also sensitive to correlations from the evolving accretion flow. 

While we have focused on classical fluctuations, the effects of multi-path propagation are also present in Hawking radiation \cite{Decanini:2011xi}, which arises via quantum mechanical fluctuations near the horizon \cite{Hawking:1974sw}. Hawking radiation itself is encoded in (quantum mechanical) correlation functions $\langle \Psi(t,\bm r) \Psi(t',\bm r') \rangle_Q$, which for a non-interacting scalar field theory satisfy the wave equation in both arguments,
\begin{align}  
\nabla^2 \langle \Psi(t,\bm r) \Psi(t',\bm r') \rangle_Q = \nabla'^2 \langle \Psi(t,\bm r) \Psi(t',\bm r') \rangle_Q = 0.
\end{align}
It is therefore reasonable to surmise that there should exist echos of Hawking radiation in $\langle \Psi(t,\bm r) \Psi(0,\bm r') \rangle_Q$.  Namely, in the coincident point limit $\bm r' \to \bm r$, $\langle \Psi(t,\bm r) \Psi(0,\bm r') \rangle_Q$ should exhibit a series of peaks at times equal to integer multiples of the photon orbit period. 
Observables that are sensitive to this correlation structure during the course of black hole evaporation
could provide much stronger evidence for Hawking radiation than the burst alone \cite{Page_Hawking_1976}. We leave a detailed study of echos in Hawking radiation for a future analysis.  

Finally, while we have analyzed the case of a scalar field near a Schwarzschild black hole, the strong suppression $\widetilde{C}_{\rm multi\mhyphen path} \sim 1/(M \omega)$ does not depend on either of these simplifications. Specifically, cancellation occurs because the phase coherence scale is comparable to the wavelength while the expected emitting region size (and, hence, the spread in multipath delay) is comparable to $M$. In contrast, the shape of the correlation function (e.g., Fig.~\ref{fig:realtimecor}) will depend on the spacetime and emission assumptions. While the coherent autocorrelation function is unlikely to be detectable for incoherent emission regions (such as synchrotron emission near a supermassive black hole), coherent emission from much smaller regions could produce detectable autocorrelation from multipath propagation. Such emission is seen in astronomical sources including pulsars and fast radio bursts.

\section{Acknowledgments}%
This work was supported by the Black Hole Initiative at Harvard University, 
which is funded by grants from the John Templeton Foundation and the Gordon and Betty Moore Foundation. We thank the National Science Foundation (AST-1716536, AST-1440254, AST-1935980, OISE-1743747) and the Gordon and Betty Moore Foundation (GBMF-5278) for financial support of this work. We thank George Wong for feedback that improved the clarity of the manuscript. 


\section*{References}
\bibliography{refs}%

\bibliographystyle{utphys}

\end{document}